\documentclass{article} 
\usepackage{spconf}
\usepackage{cite}

\usepackage{epstopdf}
\usepackage{stfloats}
\usepackage[cmex10]{amsmath}
\usepackage{amssymb}
\usepackage{graphicx}
\usepackage{multirow}
\usepackage{enumitem}
\usepackage{lipsum}
\usepackage{array}

\newcolumntype{C}[1]{>{\centering}m{#1}}
\usepackage[tight,footnotesize]{subfigure}
\usepackage{booktabs}
\usepackage{caption}
\date{}

\usepackage{enumitem}

\usepackage{url}

\usepackage{hyperref} 

\begin{document}
	
	\onecolumn
	
	\begin{description}[labelindent=0cm,leftmargin=3cm,rightmargin=3cm,style=multiline]
		
		\item[\textbf{Citation}]{M. Alfarraj, Y. Alaudah and G. AlRegib, "Content-adaptive non-parametric texture similarity measure," 2016 IEEE 18th International Workshop on Multimedia Signal Processing (MMSP), Montreal, QC, 2016, pp. 1-6.}
		
		\item[\textbf{DOI}]{\url{https://doi.org/10.1109/MMSP.2016.7813338}}
		
		\item[\textbf{Review}]{Date of publication: 21-23 Sept. 2016}
		
		\item[\textbf{Data and Codes}]{\href{https://github.com/olivesgatech/Content-Adaptive-Non-Parametric-Texture-Similarity-Measure}{\underline{GitHub Link}}}

		\item[\textbf{Bib}] {@INPROCEEDINGS\{7813338, \\
			author={M. Alfarraj and Y. Alaudah and G. AlRegib}, \\
			booktitle={2016 IEEE 18th International Workshop on Multimedia Signal Processing (MMSP)}, \\
			title={Content-adaptive non-parametric texture similarity measure},\\ 
			year={2016}, \\
			pages={1-6}, \\
			keywords={computer vision;content-based retrieval;curvelet transforms;image retrieval;image texture;singular value decomposition;visual databases;visual perception;content-adaptive nonparametric texture similarity measure;singular value decomposition;curvelet coefficients;content-based truncation;natural texture images;image perception;computer vision applications;retrieval experiment;CUReT texture databases;PerTex texture databases;Transforms;Electric variables measurement;Singular value decomposition;Computer vision;Image retrieval;Image retrieval;Image similarity;Feature extraction;Texture analysis}, \\
			doi={10.1109/MMSP.2016.7813338}, \\
			ISSN={2473-3628}, \\
			month={Sept},\}}

		\item[\textbf{Copyright}]{\textcopyright 2018 IEEE. Personal use of this material is permitted. Permission from IEEE must be obtained for all other uses, in any current or future media, including reprinting/republishing this material for advertising or promotional purposes,
			creating new collective works, for resale or redistribution to servers or lists, or reuse of any copyrighted component
			of this work in other works. }
		
		\item[\textbf{Contact}]{\href{mailto:motaz@gatech.edu}{motaz@gatech.edu}  OR \href{mailto:alregib@gatech.edu}{alregib@gatech.edu}\\ \url{http://ghassanalregib.com/} \\ }
	\end{description}
	
	\thispagestyle{empty}
	\newpage
	\clearpage
	\setcounter{page}{1}
	
	\twocolumn
	
\title{Content-adaptive Non-parametric texture similarity measure}
\name{Motaz Alfarraj, Yazeed Alaudah, and Ghassan AlRegib}
\address{Center for Signal and Information Processing (CSIP) \\ School of Electrical and Computer Engineering \\ Georgia Institute of Technology \\ \{motaz,alaudah,alregib\}@gatech.edu}

\maketitle

\begin{abstract}
In this paper, we introduce a non-parametric texture similarity measure based on the singular value decomposition of the curvelet coefficients followed by a content-based truncation of the singular values. This measure focuses on images with repeating structures and directional content such as those found in natural texture images. Such textural content is critical for image perception and its similarity plays a vital role in various computer vision applications. In this paper, we evaluate the effectiveness of the proposed measure using a retrieval experiment. The proposed measure outperforms the state-of-the-art texture similarity metrics on \texttt{CUReT} and \texttt{PerTex} texture databases, respectively.
\end{abstract}

\section{Introduction}
\label{into}
The exponential growth of visual and pictorial content undoubtedly drives an increasing need for image similarity quantification that can be utilized for various computer vision applications. The similarity of two images is often measured with respect to some attributes, for instance, shape, color or texture. The attributes to be utilized in the measure are application-dependent. For example, texture attributes are widely used in constructing images depicting natural scenes for virtual reality environments in what is known as texture synthesis. In example-based texture synthesis, texture images are generated or extended in a way such that they have the same textural feel and pattern as the example texture image without naively copying the example image \cite{synthesisReview}. The synthesis process can be optimized by maximizing the textural similarity between the example and synthesized images \cite{SimilaritySynthesis}.

Furthermore, the proposed measure utilizes the curvelet transform \cite{candes05} to exploit the content of textured images with a granularity that makes the resulting feature vector effective in capturing the subtle differences between images. Due to its effective capturing of the local structure differences, the proposed measure can also be used in a two-tier quality assessment mechanism where the first-tier is a specialized global quality measure such as the ones proposed in \cite{temel2015persim}\cite{fsim}.

In this paper we propose a texture similarity metric based on the singular value decomposition (SVD) of the curvelet coefficients in which the singular values are trimmed adaptively via effective rank approximation. A brief description of the curvelet transform is presented in Section \ref{curvelet}. The methodology is outlined in Section \ref{method}. Section \ref{databases} provides details about the databases used in the experiments. In Section \ref{experiments}, we evaluate and compare the performance of the proposed metric to the performance of common similarity metrics and texture similarity measures in the literature. A discussion and analysis of the results follows in Section \ref{discussion}. 

\section{Background}
\subsection{Texture Similarity}
\begin{figure}[h]
\centering
\subfigure[Image 1]{
\includegraphics[width = 0.3\columnwidth]{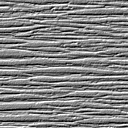}}
\subfigure[Image 2]{
\includegraphics[width = 0.3\columnwidth]{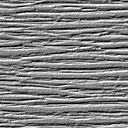}}
\subfigure[absolute difference]{
\includegraphics[width = 0.3\columnwidth]{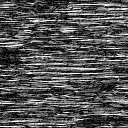}}
\caption{Two images with similar texture and their absolute difference.}
\label{fig:textureDiff}
\end{figure}

Texture-based image similarity or distance measures differ from their generic counterpart measures such as Peak Signal-to-Noise Ratio (PSNR) and Mean Square Error (MSE) in that they capture the content of an image rather than assuming a pixel-to-pixel correspondence. Figure \ref{fig:textureDiff} shows an example of two images with the same texture and their absolute difference. This examples shows that pixel-based comparisons are not fit for measuring texture similarity. Structural Similarity Metric (SSIM) improves upon pixel-to-pixel metrics by capturing structure using low-level local statistics in the spatial domain (S-SSIM)\cite{ssim} or the complex wavelet domain (CW-SSIM)\cite{cwssimJournal}. Other metrics have been proposed to measure texture similarity in the spatial domain such as Local Binary Patterns (LBP) \cite{LBP} which characterizes the texture by constructing binary pattern maps based on the value of a pixel relative to its neighbors and obtaining statistics from the histograms of these patterns for different radii. Structural Texture Similarity Metric (STSIM)\cite{stsim} uses subband statistics and correlations in a multiscale frequency decomposition, namely steerable pyramid. Low-level statistics of the curvelet coefficients were used in the development of a content based image retrieval (CBIR) metric for texture images \cite{curveletCBIR}. In general, the practice of using a filter bank followed by statistics extraction or energy pooling is very common in texture image analysis and has proven effective for such applications \cite{waveletsvd}\cite{curveletCo}\cite{curveletCharacteriation}\cite{KLDwavelet}.\\

\subsection{The Curvelet Transform}\label{curvelet}

\begin{figure}[ht!]
\centering
\subfigure[Frequency viewpoint]{
\includegraphics[width = 0.4\linewidth]{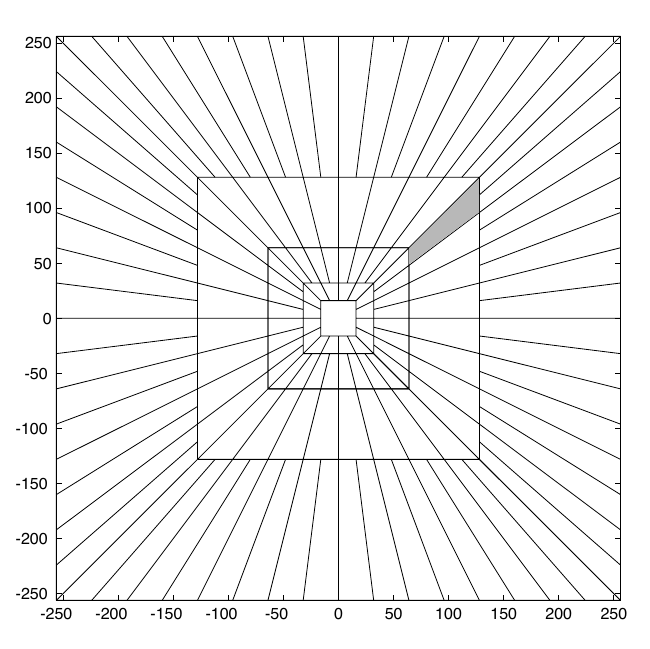}
\label{fig:curveletFreq}}
\quad
\subfigure[Spatial viewpoint]{
\includegraphics[width = 0.4\linewidth]{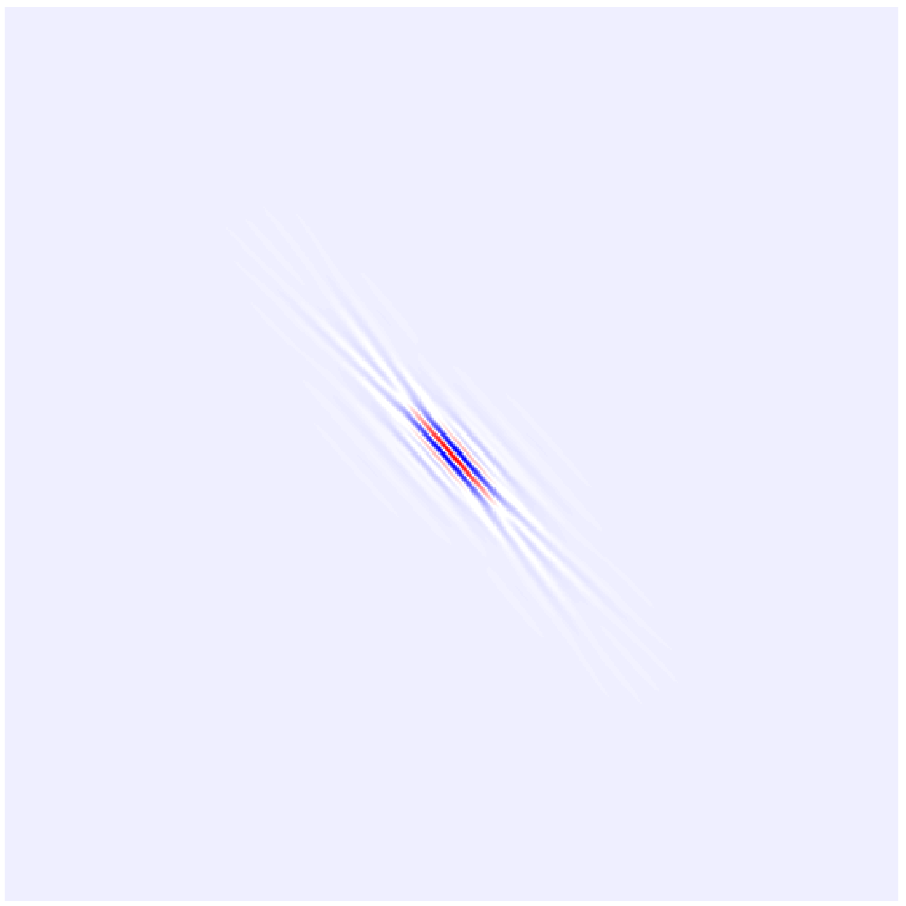}
\label{fig:curveletSpatial}}
\caption{Frequency and spatial viewpoints of a curvelet wedge. Adapted from \cite{candes05}.}
\label{fig:curvelet}
\end{figure}

The curvelet transform is a directional multiscale decomposition \cite{candes05} that provides an efficient way to represent images with high directional content. It has been shown  that the curvelet transform provides an optimal sparse representation for curve-like structures or edge-like phenomena when compared to wavelets~\cite{c2curve}. The curvelet coefficients are obtained by tiling the spectrum of the image at different scales and directions as shown in Figure \ref{fig:curveletFreq}. Then, each wedge is transformed back to the spatial domain by taking the 2D inverse Fourier transform after applying a smoothing function and wrapping it around the origin to fit the trapezoid-shaped tile into a rectangle. A wedge in the frequency domain is represented by a needle shape in the spatial domain with a direction perpendicular to the orientation of wedge and with a width inversely proportional to the scale number. Figure \ref{fig:curveletSpatial} shows the spatial representation of the highlighted wedge in Figure \ref{fig:curveletFreq}.

\section{METHODOLOGY} \label{method}

\begin{figure}[ht!]
\centering
\includegraphics[width = \columnwidth]{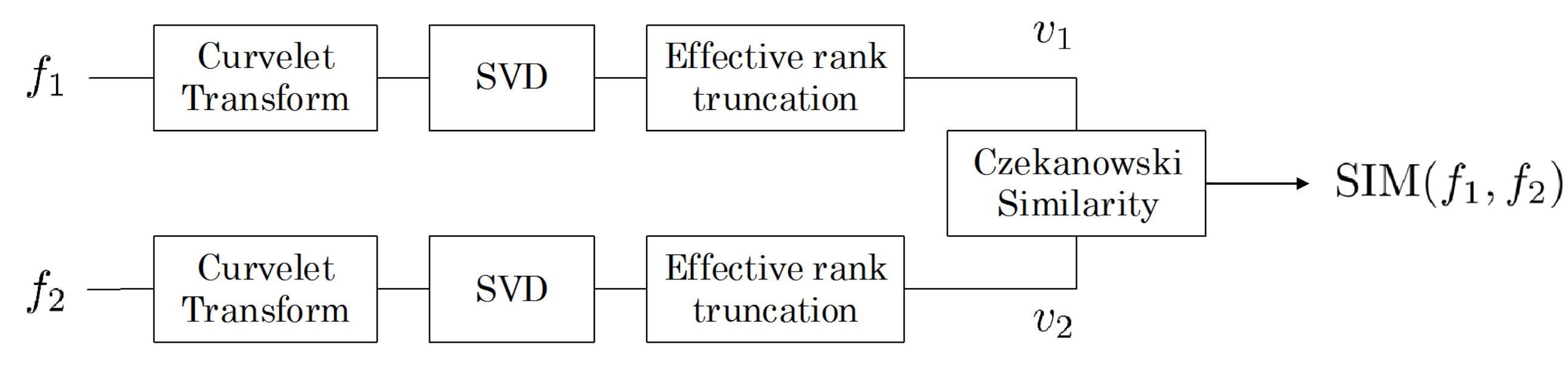}
\caption{Block diagram of the proposed method}
\label{fig:block_diagram}
\end{figure}

The proposed measure calculates the similarity score between two images based on the similarity between their corresponding feature vectors. The feature vector of a given image is a collection of all effective singular values for all scales and orientations of the curvelet coefficients of an image. The block diagram of the process is depicted in Figure \ref{fig:block_diagram}. 

For a grayscale image, $f$, the curvelet coefficients are obtained for all scales, $j=\{1,2,\dots,J\}$, and orientations, $k =\{1,2, \dots, K(j)\}$, where $J$ is the total number of scales and $K(j)$ is the number of orientations for scale $j$. However, due to the conjugate symmetry of the Fourier coefficients for real images, we will consider half the number of orientations. Note that the innermost and outermost scales are not split into orientations. Then, the singular values are calculated as $\bar{\sigma}_{[j,k]} = [\sigma_1,\dots,\sigma_L]$ where $\sigma_1 \geq \sigma_2 \hdots \geq \sigma_L$ and $L$ is the smallest dimension of the coefficients matrix. Ideally, if the rank of a matrix is $p$, only the first $p$ singular values are non-zero and the remaining ones are identically equal to zero. However, when we consider SVD on digital images that are subject to different types of noise, the number of non-zero singular values is greater than $p$. In most cases, none of the singular values are identically zero even for a rank-deficient matrix. Roy and Vetterli \cite{roy2007effective} proposed the effective rank as a method to estimate the actual rank by estimating the effective dimensionality of a matrix. To calculate the effective rank, a singular value distribution is defined as: \begin{equation}\label{SVdistribution}
p_k = \frac{\sigma_k}{\|\bar{\sigma}\|_1} \text{for } k = 1,\dots, L,
\end{equation}
where $\|\cdot \|_1$ is the $\ell_1$ norm. Then, the effective rank is calculated as a function of the entropy of a signal with the singular value distribution defined in Eq.~\ref{SVdistribution},
\begin{equation}\label{eq:effectiveRank}
\text{EffectiveRank} = \exp{\left(-\sum_{k=1}^{L} p_k \log p_k\right)},
\end{equation}
resulting in a real number less than or equal to $L$ with equality if and only if all singular values are equal. We use $\log$ to denote the natural logarithm and we follow the convention that $0\log 0 = 0$. 

For each set of coefficients, the effective rank $q$ is calculated as in Eq.~\ref{eq:effectiveRank}. A new vector of ``effective'' singular values is formed by keeping the first $\lfloor q \rfloor$ singular values, where $\lfloor q \rfloor$ is the integer part of $q$. The remaining singular values are set to 0, i.e. for scale $j$ and orientation $k$, we form the vector $\hat{\sigma}_{[j,k]} = [\sigma_1,\dots,\sigma_q,0,\dots,0]$. The feature vector of $f$ is obtained by concatenating all $\hat{\sigma}_{[j,k]}$ for all scales and half the number of orientations,
\begin{equation}\label{eq:featVec}
\bar{v} = [\hat{\sigma}_{[1,1]},\hat{\sigma}_{[2,1]}, \dots, \hat{\sigma}_{[2,K(2)/2]},\hat{\sigma}_{[3,1]}\dots,\hat{\sigma}_{[J,1]}].
\end{equation}

As we described earlier, the smaller singular values can be induced by noise, and minimizing the effect of such values is desirable. However, in the proposed method, the non-effective singular values are filtered out according to the effective rank. In order to compare singular values of relatively equal significance but different magnitudes, we propose using an $\ell_1$-norm based metric, namely, the Czekanowski similarity. The similarity between two images, $f_1$ and $f_2$, is found as the Czekanowski similarity coefficient between their corresponding feature vectors $\bar{v}_1$ and $\bar{v}_2$ as: 

\begin{equation}
\text{SIM} (f_1,f_2) = 1-\frac{\|\bar{v}_1-\bar{v}_2\|_1}{\|\bar{v}_1+\bar{v}_2\|_1}.
\end{equation}
Since the singular values are non-negative by definition, $\text{SIM}(f_1,f_2) \in [0,1]$ with scores closer to $1$ indicating higher similarity.
\section{Databases}
\label{databases}

The proposed measure has been tested on two different texture databases and one seismic database. We use $C$ and $S$ to indicate the number of classes and the number samples per class, respectively. Sample patches from each database are shown in Figures \ref{fig:SamplesCUReT} and \ref{fig:SamplesPerTex}. The databases used in the experiments are:
 
\begin{enumerate}[nolistsep]
\item \texttt{CUReT} \cite{curet}: Non-overlapping patches of size $128\times 128$ were cropped from the central $256\times 256$ pixels of all images with viewing condition number 55. Please refer to \cite{curet} for details on the viewing conditions. ($C=61$, $S=3$).   

\begin{figure}[h]
\centering
\includegraphics[width = \columnwidth]{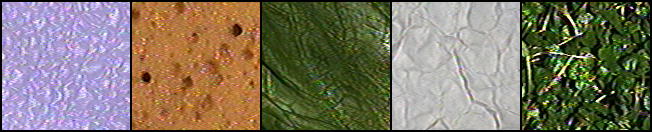}
\caption{Sample images from CUReT database.}
\label{fig:SamplesCUReT}
\end{figure}

\item \texttt{PerTex}\cite{PerTex}: Each image in the database was downsampled by a factor of $4$ then divided into 4 non-overlapping quadrants of size $128\times 128$. ($C=334,S=4$).

\begin{figure}[h]
\centering
\includegraphics[width = \columnwidth]{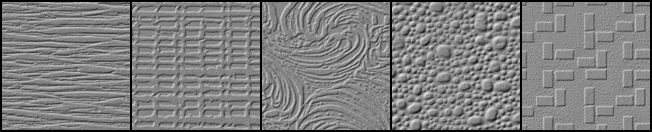}
\caption{Sample images from PerTex database.}
\label{fig:SamplesPerTex}
\end{figure}

\end{enumerate}

\section{Performance Evaluation}\label{experiments}
A retrieval experiment was set up to evaluate the performance of the proposed measure on each of the four databases. Note that for color images, only the luminance channel was used in the experiments.  Each image, $f_i$, in a dataset of $C\times S$ images was compared to all other $(C\times S-1)$ images. The images are then ranked according to their similarity (or distance) and $(S-1)$ images, $g_i^1, \dots g_i^{S-1}$, with the highest similarity (or lowest distance) are retrieved. The superscript denotes the rank of the retrieved image. 

\subsection{Retrieval Metrics}
To quantify the goodness of a similarity measure, the following information retrieval metrics were used:
\begin{itemize}

\item Precision at One (P@1) is defined as the percentage of the first retrieved images that are relevant to the query image (i.e. they belong to the same class as that of the query image),
\begin{equation}\label{P@1}
\text{P@1} = \frac{\sum_{i=1}^{CS} \mathbf{1}_{\{T: f_i\in T\}}(g_i^1)}{CS}.
\end{equation} 
where $\mathbf{1}_{\{T: f_i\in T\}}(g_i^j)$ is the indicator function that is equal to 1 when $f_i$ and $g_i^j$ are of the same class $T$, and is equal to zero otherwise.

\item Mean reciprocal rank (MRR) is the average reciprocal rank of the first correctly retrieved image, 
\begin{equation}\label{MRR}
\text{MRR} = \frac{1}{CS} \sum_{i=1}^{CS} \frac{1}{\text{rank}_i(1)}, 
\end{equation} 
where $\text{rank}_i(m)$ is the rank of the $m^\text{th}$ correctly retrieved image for a query image $f_i$. 

\item Mean Average Precision (MAP) is typically used when there is more than one relevant image for a given query image. It takes into account the order in which the images were retrieved. Average precision (AP) is calculated for each query instance as: 

\begin{equation}\label{AP}
\text{AP}(i) = \frac{1}{S-1} \sum_{m=1}^{S-1} \frac{m}{\text{rank}_i(m)}.
\end{equation}
Then,
\begin{equation}\label{MAP}
\text{MAP} = \frac{1}{SC}\sum_{i=1}^{SC}\text{AP}(i).
\end{equation} 
\end{itemize}

All of the used information retrieval metrics are in the range $[0,1]$ with $1$ being a perfect score. The results are summarized in Table \ref{table:results}. 

\subsection{Receiver Operating Characteristic Curve}
\begin{figure}[ht!]
\centering 
\subfigure[]{\includegraphics[width = 0.9\linewidth]{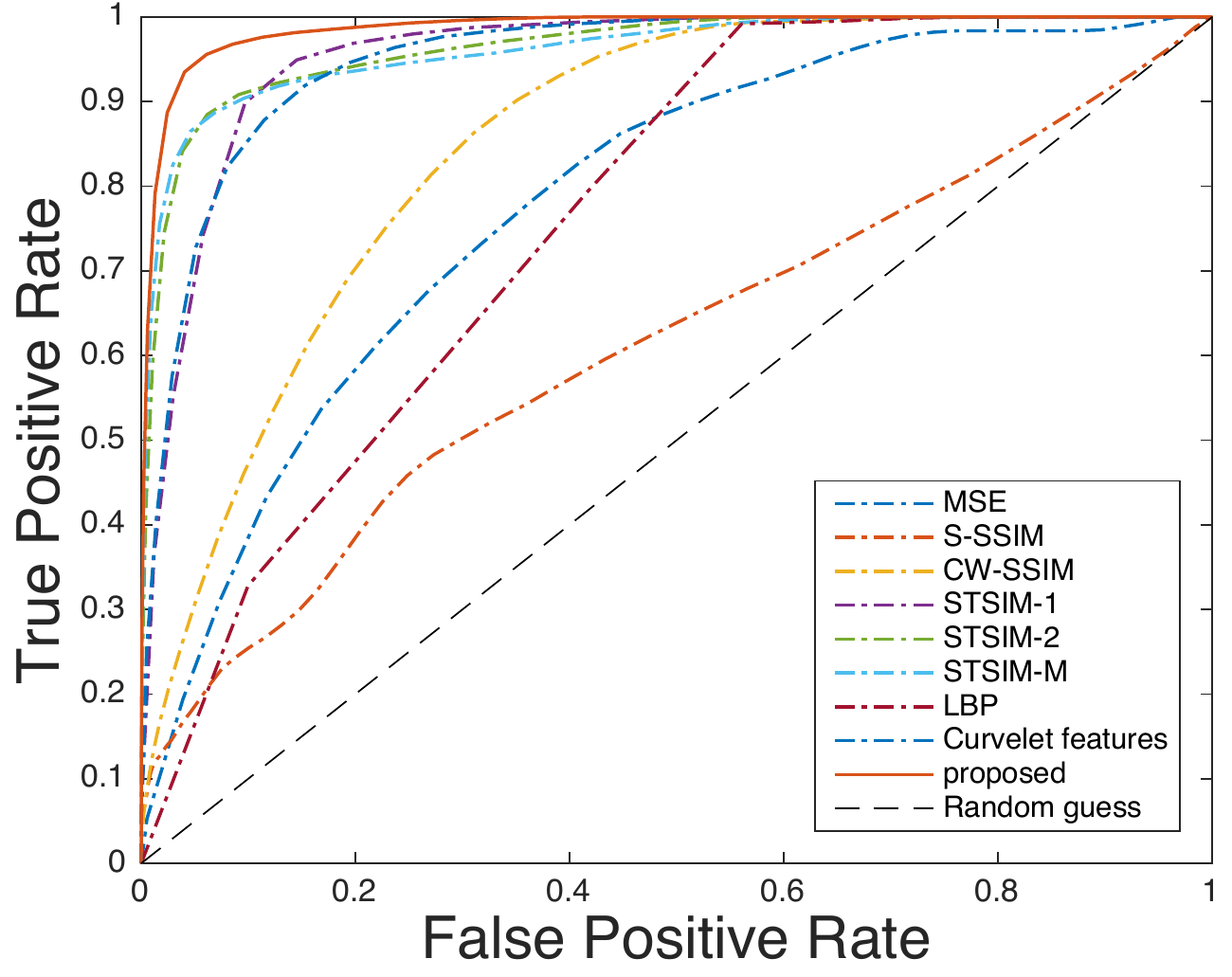}}
\subfigure[]{\includegraphics[width =0.9 \linewidth]{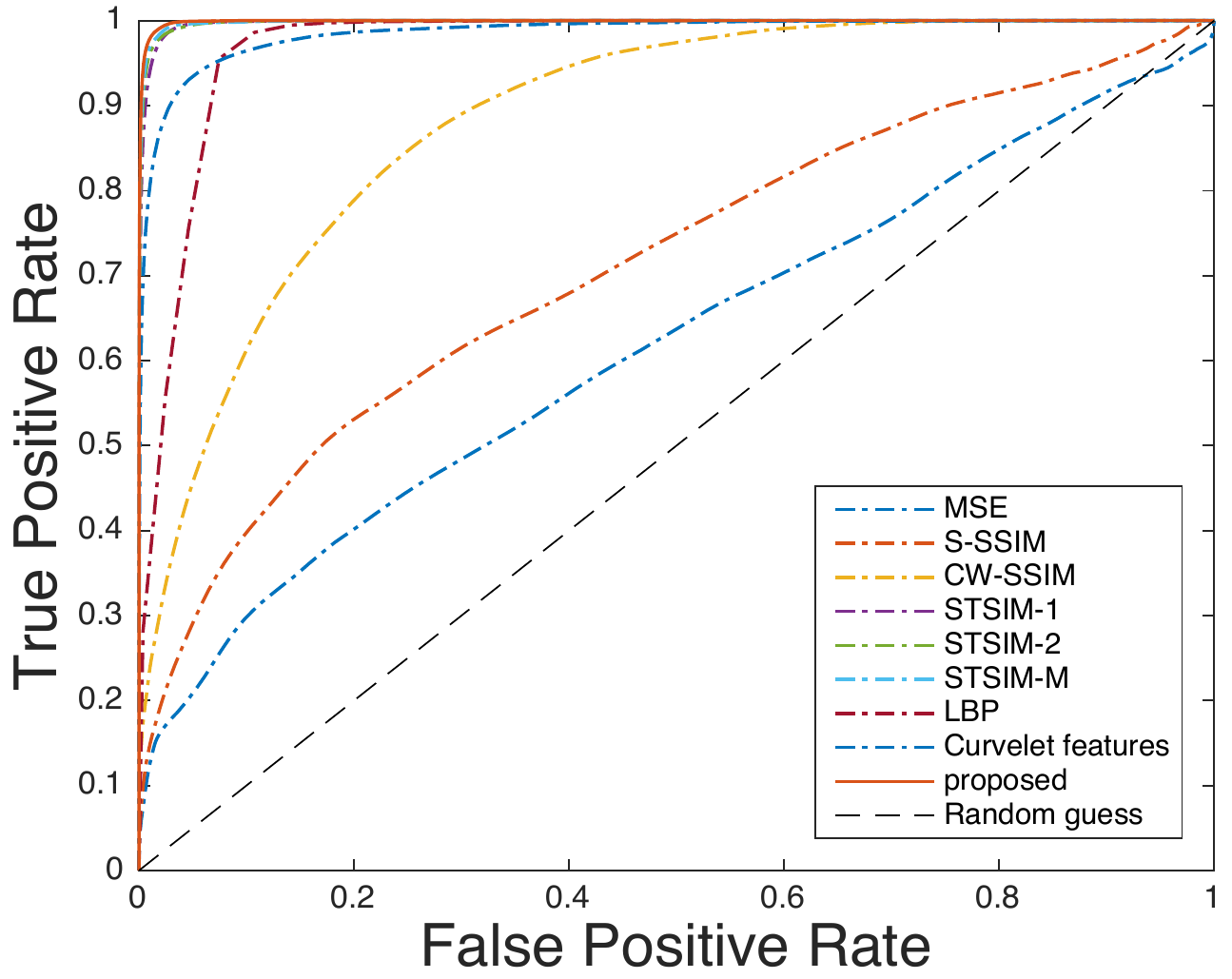}} 
\caption{ROC curves for different similarity/distance measures on (a) \texttt{CUReT} and (b) \texttt{PerTex} databases.}
\label{fig:ROC}
\end{figure}

The retrieval experiment can been seen as a binary classification problem where a binary decision (similar/dissimilar) is made for each retrieval instance. The performance of the discriminative power of such classifier can be illustrated using the Receiver Operating Characteristic (ROC) curve. The ROC curve is a plot of the True Positive Rate (TPR) for different threshold on the False Positive Rate (FPR). The area under the curve (AUC) can be used as a performance measure. The ideal ROC curve is the one where TPR is equal to 1 for all values of FPR. Thus, an ROC curve that is pushed towards to the upper left corner reflects a good performance or, equivalently, AUC is closer to 1.  The ROC curves for different similarity measures and their corresponding AUC are presented in Figure \ref{fig:ROC} and Table \ref{table:AUC}, respectively, on the two databases described in section \ref{databases}. 

The following similarity/distance measures were used in the experiments:
\begin{enumerate}[nolistsep]
\item Mean squared error (MSE). 
\item S-SSIM with default parameters \cite{ssim}. 
\item CW-SSIM with 4 scales and 8 orientations \cite{cwssimJournal}. 
\item STSIM-1,STSIM-2 and STSIM-M with best results given when parameters are set to 3 scales and 4 orientations \cite{stsim}. 
\item LBP with radii of 8 and 24 \cite{LBP}. 
\item $\ell_2$ norm on Curvelet Features \cite{curveletCBIR}.
\end{enumerate}
 
 The codes for S-SSIM, CW-SSIM, LBP distance were obtained from their respective authors. All other metrics were implemented as presented in the cited papers.

\begin{table}
\centering
\resizebox{\linewidth}{!}{
	\begin{tabular}{l || c c c | c c c }
		\toprule
		Database &\multicolumn{3}{c|}{\texttt{CUReT}}&\multicolumn{3}{c}{\texttt{PerTex}}\\ 
		\toprule
		Metric				&P@1&MRR&MAP		&P@1&MRR&MAP\\\midrule
		
		MSE					&0.1093 & 0.1867 & 0.1728	 
		&0.1115 & 0.1296 & 0.0641 	 \\
		
		S-SSIM	\cite{ssim}			&0.0546 & 0.0952 & 0.0935  		
		&0.1572 & 0.1855 & 0.1051 	\\
		
		CW-SSIM	 \cite{cwssimJournal}			&0.1366 & 0.2638 & 0.1967	
		&0.1826 & 0.2527 & 0.1665  \\

		STSIM-1 \cite{stsim}				&0.9071 & 0.9447 & 0.9048 	
		&0.9513 & 0.9658 & 0.9155 	  \\
		
		STSIM-2 \cite{stsim}				&0.8852  & 0.9248 & 0.8500 		
		&0.9401 & 0.9627 & 0.9152     \\
		
		STSIM-M \cite{stsim}				&0.8798 & 0.9170 & 0.8516 	
		&0.9731 & 0.9845 & 0.9610    \\										
		
		LBP		\cite{LBP}			&0.8415 & 0.8843 & 0.8347 	
		&0.9117 & 0.9407 & 0.8628  \\				
		
		Curvelet features \cite{curveletCBIR}	&0.8142 & 0.8649 & 0.7669  	
		&0.9499 & 0.9658 & 0.9098  \\
		
		Proposed		&\textbf{0.9617}&\textbf{0.9732}&\textbf{0.9304}
		
		&\textbf{0.9880}&\textbf{0.9917}&\textbf{0.9736}\\
		\bottomrule
\end{tabular}}

\caption{Performance evaluation on different databases}
\label{table:results}
\end{table}

\begin{table}[ht!]
\centering

\begin{tabular}{l || c | c }
	\toprule
	Database &\texttt{CUReT}&\texttt{PerTex}\\ 
	\midrule
	Metric &AUC&AUC\\ 
	\toprule
	MSE										&0.6168 & 0.7926 \\					
	S-SSIM	\cite{ssim}						&0.7119 & 0.6109 \\			
	CW-SSIM	 \cite{cwssimJournal}			&0.8849 & 0.8563 \\
	STSIM-1 \cite{stsim}					&0.9978 & 0.9750 \\
	STSIM-2 \cite{stsim}					&0.9981  & 0.9703 \\
	STSIM-M \cite{stsim}					&0.9985 & 0.9655 \\								
	LBP		\cite{LBP}						&0.9891 & 0.8432 \\				
	Curvelet features \cite{curveletCBIR}	&0.9864 & 0.9618 \\
	Proposed									&\textbf{0.9991}&\textbf{0.9907}\\
	\bottomrule
\end{tabular}

\caption{Performance evaluation on different databases}
\label{table:AUC}
\end{table}

\begin{figure*}[ht!]
\centering 
\subfigure[CUReT]{\includegraphics[width =\linewidth]{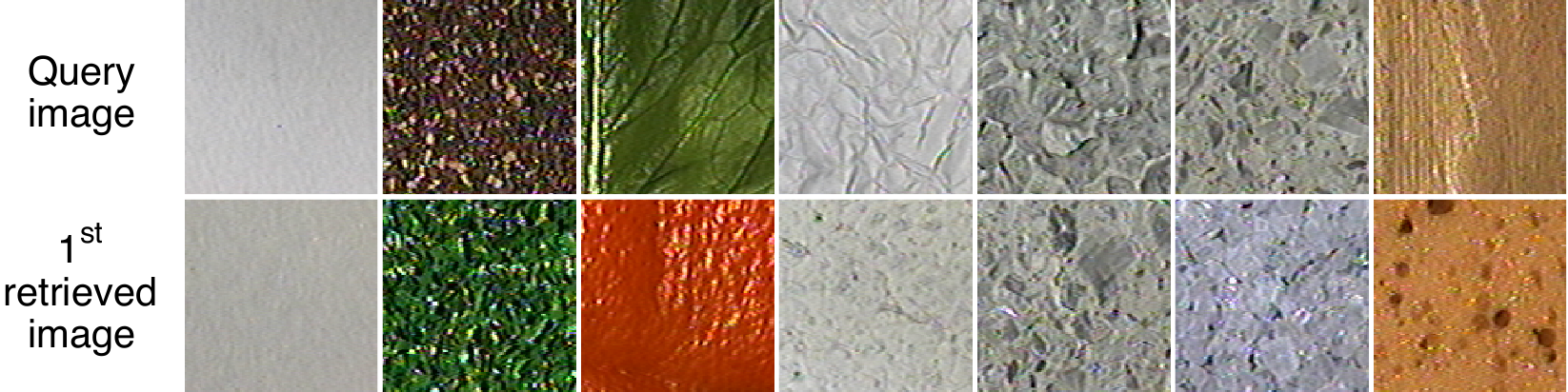}}
\subfigure[PerTex]{\includegraphics[width =\linewidth]{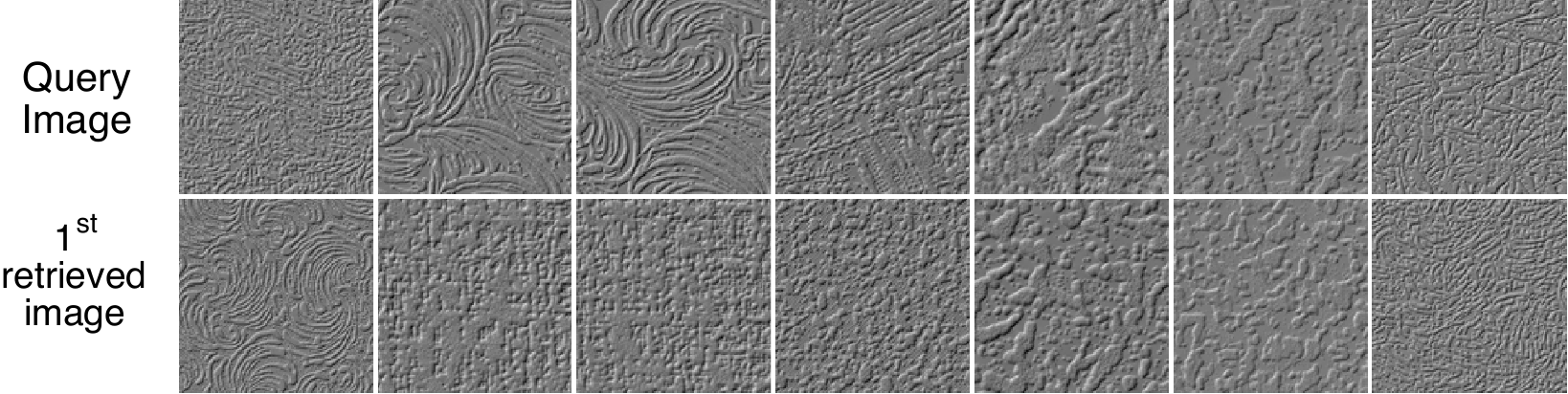}}
\caption{Examples of first retrieved image that are irrelevant to the query image from(a) \texttt{CUReT} and (b) \texttt{PerTex} databases.}
\label{fig:retrieved}
\end{figure*}

\section{Discussion}
\label{discussion}
Table \ref{table:results} shows retrieval performance for the different measures discussed in this paper. The power of the measures that are specially designed for texture images is clear from numbers in the table compared to the generic similarity (or distance) measures. Furthermore, the proposed measure outperforms all other measures on the two texture databases in all retrieval metrics. Even though the databases are very different in terms of the type of image they have, the proposed measure maintains a consistent performance. \texttt{CUReT} is a databases of color images and we have used the luminance channel only in the evaluation. Some images in this databases have very similar textures and are only distinguishable by color. Figure \ref{fig:retrieved} shows some examples where the first retrieved images are not relevant to the query image. Although the retrieved images on CUReT database are not of the same class as that of the query image, they have very similar texture patterns making them hard to distinguish using structure cues only. Furthermore, PerTex database has some classes that are distributable on smaller windows (cropped images) as shown in some of the examples in Figure \ref{fig:retrieved}. Furthermore, ROC curve for the proposed measure shows its ability to effectively distinguish different classes even for very tight constraints on the False Positive Rate threshold as illustrated in Figure \ref{fig:ROC} and Table \ref{table:AUC}.

\section{Conclusion}
\label{conclusion}
In this paper we proposed a non-parametric textural similarity measure for texture images based on the effective singular values of the curvelet coefficients. For each set of curvelet coefficients, the number of effective singular values is adaptively determined by the effective rank of their distribution. The similarity of two images is calculated as the Czekanowski coefficient between the corresponding vectors of their effective singular values. The measure was tested on two texture databases as well as a seismic database and it tops the list of all similarity and distance measures with a consistent performance on all databases used in the experiments.

\section{Acknowledgment}
This work is supported by the Center for Energy and Geo Processing (CeGP) at Georgia Tech and King Fahd University of Petroleum and Minerals (KFUPM).



\end{document}